\documentstyle[12pt]{article}

\begin{document}

\title{Possible candidate of $0^+$ $s\bar{s}s\bar{s}$ state }
\author{Bing An Li\\
University of Kentucky\\
Lexington, KY 40506, USA}

\maketitle

\begin{abstract}
The possibility of the $0^+$ $\eta\eta$ resonance $f_0(2100)$ as a candidate of the $Q^2\bar{Q}^2$ state $C^{ss}(36)$ is explored.
The $\eta\eta$ channel of $f_0(2100)$ is the dominant decay mode, $\eta\eta'$ channel has less decay rate, the decay rate of the $\eta'\eta'$ channel 
is very small. The $\pi\pi,\;K\bar{K},\;4\pi$ modes are at next leading order in $N_C$ expansion. Other possible decay modes are discussed. 
\end{abstract}
\newpage

Among many $I^G(J^{PC}) = 0^+ (0^{++})$ resonances discovered [1] around 2 $\textrm{GeV}$ there are $f_0(2100)$ and $f_0(2020)$ two scalar mesons [1].
The parameters of the $f_0(2100)$ are determined to be [1] 
\[M = 2103 \pm 8 \textrm{MeV},\;\; \Gamma = 209 \pm 19 \textrm {MeV}.\]
The $f_0(2020)$ is a broad scalar meson whose decay width is about 400 $\textrm{MeV}$ and it has many decay modes: $\rho\pi\pi,\;\pi^0\pi^0,\;\rho\rho,\;
\omega\omega,\;\eta\eta,...$ [1]. The $f_0(2100)$ is different from the $f_0(2020)$. In this paper the possible nature of the $f_0(2100)$ is investigated.
In 1993 $\eta\eta$ resonances have been reported in following processes [2]
\[ \bar{p} + p \rightarrow 3\pi^0,\;2\pi^2,\;\pi^0 2\eta,\;3\eta.\]
The $f_0(2100)$ is one of the three $\eta\eta$ resonances and its mass and decay width are determined to be
\[M = 2104 \pm 20 \textrm{MeV},\;\Gamma = 203 \pm 10 \textrm{MeV}.\]
This state has not been identified definitely in the $\pi\pi$ channel
and the quantum numbers are not determined in this study. 
In Ref. [3] the $f_0(2100)$ has been discovered in 
\[ \bar{p} + p \rightarrow \eta \eta,\;\eta \eta',\; ....\]
The $f_0(2100)$ appears strongly in $\eta\eta$ channel and 
\[ M = 2105 \pm 10 \textrm{MeV},\; \Gamma = 200 \pm 25 \textrm{MeV}.\]
The $f_0(2100)$ appears weakly in the $\pi^0 \pi^0 $ data, contributing only $(4.6 \pm 1.5)\%$
of the cross section, comparing with $(38 \pm 5)\%$ in $\eta\eta$ channel. 
The cross section for $\eta\eta'$ contains a weak peak at about 2150 $\textrm{MeV}$.
 
Recently, the BES III Collaboration has reported the discovery of the $\eta\eta$ resonance $f_0(2100)$ in $J/\psi\rightarrow\gamma\eta\eta$ [4].
Its mass and decay width are determined to be
\[m = 2081 \pm 13 \textrm{MeV},\;\Gamma = 273^{+27}_{-24} \textrm{MeV}\]
respectively, which are in agreement with the measurements [2,3].
The product branching ratio is measured to be
\[B(J/\psi\rightarrow\gamma f_0(2100)\rightarrow\gamma\eta\eta) = (1.13^{+0.09}_{-0.1})\times 10^{-4}.\]

On the other hand, the discovery of $f_0(2100)$ has not been reported in $K\bar{K}$ channels of $J/\psi\rightarrow \gamma K\bar{K}$ [4]
and $p p\rightarrow p_f (K\bar{K}) p_s$ [5,6]. 

In Ref. [6] in $J/\psi\rightarrow \gamma\pi^+\pi^-$ besides a $2^{++}$ state $\theta(1700)$ a X(2100) is reported 
\[M = 2027 \pm 12 \textrm{MeV},\; \Gamma = 220 \pm 30 \textrm{MeV} .\] 
However, it is claimed that the angular distributions of X(2100) are similar to those of the $\theta(1700)$ which has been determined to be a $2^{++}$ state.
In Ref. [7] in the decay $J/\psi\rightarrow \gamma\pi\pi$ a wide resonance $f_0(2020)$ is seen, which is listed in Ref. [1]. 
It seems that the results from these two experiments do not agree each other. 
On the other hand, in Ref. [1] the scalar resonance $f_0(2100)$ is not listed in the $\pi\pi$ channel in $J/\psi\rightarrow \gamma\pi\pi$.

The experimental study of the scalar resonance in the 4$\pi$ channel lasts a pretty long time. 
Now it is needed to check whether the $f_0(2100)$ has $4\pi$ decay mode.
$J/\psi\rightarrow\gamma 4\pi$ and $p p\rightarrow p_f(4\pi)p_s$ are the two processes to search for $X\rightarrow 4\pi$.
\begin{enumerate}
\item MARK II has done a study of $J/\psi\rightarrow \gamma 4\pi$ [8] and it is found that the $\gamma\rho\rho$ are the components of these channels.
      It is no mentioned whether there is resonance of $\rho\rho$ around 2 $\textrm{GeV}$. 
\item In Ref. [9] a $0^{-+}$ resonance $\eta(2100)$ which decays to both $\rho^+\rho^-$ and 
      $\rho^0\rho^0$ has been observed by DM2 and $f_0(2100)\rightarrow\rho\rho$ has not been reported. 
\item MARK III did a study on $J/\psi\rightarrow\gamma \pi^+ \pi^- \pi^+ \pi^-$ [10]. The $f_0(2104)$ has been reported in the $4\pi$ channel with 
      $\Gamma = 203 \textrm{MeV}$ and $\sigma\sigma$ is the dominant decay channel. 
      Large branching ratios are reported 
      \[B(J/\psi\rightarrow\gamma X(2104))B(X(2104)\rightarrow4\pi) = (3.0 \pm 0.8)\times 10^{-4},\]
      \[B(J/\psi\rightarrow\gamma X(2104))B(X(2104)\rightarrow\rho\rho) = (6.8 \pm 1.8)\times 10^{-4}.\]
\item BES has reported a partial wave analysis of $J/\psi\rightarrow\gamma \pi^+ \pi^- \pi^+ \pi^-$ [11]. $f_0(2100)$ has been found in the $4\pi$ channel with
      \[M = 2090 ^{+30}_{-30} \textrm{MeV},\;\Gamma = 330^{+100}_{-100}  \textrm{MeV}.\]
      $f_0(2100)\rightarrow\sigma\sigma$ decay is reported. From the values of the mass and width of $f_0$ is hard to say the resonance is $f_0(2100)$ or $f_0(2020)$.
\end{enumerate}
From the current data it is difficult to draw a conclusion whether the $f_0(2100)$ has been found in the 4$\pi$ channel produced in the $J/\psi$
radiative decays.

The process $p p\rightarrow p_f(4\pi)p_s$ has been studied by WA102 [12] and broad scalar with
\[m = 2020 \pm 35 \textrm{MeV},\;\Gamma = 410 \pm 50 \textrm{MeV}\]
has been found. The resonance decays to $\rho\pi\pi$ and $\rho\rho$.
It is the $f_0(2020)$ resonance.       
In Ref. [13] a spin analysis of the $4\pi$ channels produced in central pp interactions has been done. It is mentioned that the $J^P = 0^+$ $\rho\rho$ 
distribution shows a peak at 1.45 $\textrm{MeV}$ together with a broad enhancement around 2 $\textrm{GeV}$.
These experiments show that the $0^+$ resonance found in the 4$\pi$ channel produced in pp collision is the broad $f_0(2020)$.

The experimental data mentioned above show that the scalar resonance $f_0(2100)$ discovered in $p \bar{p}$ annihilation and $J/\psi$ radiative decay decays
to $\eta\eta$ dominantly. It has weak coupling with $\pi\pi$ channel in $p \bar{p}$ annihilation and it is not found in $J/\psi\rightarrow \gamma\pi\pi$.
The $f_0(2100)$ weakly decays to $\eta\eta'$. It is not discovered in the $K\bar{K}$ channels of $J/\psi$ radiative decay and $p \bar{p}$ collision.
The data of $p \bar{p}\rightarrow p_s (4\pi) p_f$ show that there is no sign of the $f_0(2100)\rightarrow 4\pi$.

The authors of Ref. [3] claim that 
$f_0(2100)$ decays dominantly through a $s\bar{s}$ component and the strong production in $p\bar{p}$ strongly suggest 
exotic character. It is either a glueball or a hybrid and there maybe mixing with $q\bar{q}$ and $s\bar{s}$.  

In this paper possible exotic character of the $f_0(2100)$ is investigated. We explore the possibility that the $f_0(2100)$ is
a four quark state of $s\bar{s}s\bar{s}$ [14]. 
The possible direct decay channels of $s\bar{s}s\bar{s}$ are $\eta\eta,\;\eta\eta',\;\eta'\eta',\;\phi\phi,$ etc..
The $\eta\eta$ mode has the largest phase space. Therefore, the $\eta\eta$ is the dominant decay mode of the $f_0(2100)$. 
The $\pi\pi,\;K\bar{K},\;4\pi$ decays of the $f_0(2100)$ are via the meson loop diagrams, $\eta\eta\rightarrow \pi\pi,\;K\bar{K},\;4\pi$ completed.
In a meson theory [15] it shows that the tree diagrams of mesons are at the leading order in the $N_C$ expansion and the 
meson loop diagrams are at next leading order. 
For example, in this meson theory [15] the amplitude of $\phi\rightarrow K\bar{K}$ is at $O(N_C)$. The decay $\phi\rightarrow\rho\pi$ is via one-loop meson diagrams
completed. The amplitude of this decay is at O(1) in the $N_C$ expansion.   
Comparing with $\phi\rightarrow K\bar{K}$, the decay $\phi\rightarrow\rho\pi$ is suppressed.
For the decays of the $f_0(2100)$ the $f_0(2100)\rightarrow\eta\eta$ is resulted in tree diagram and 
the $\pi\pi,\;K\bar{K},\;4\pi$ channels are resulted in loop diagrams of mesons, therefore, they are suppressed in the $N_C$ expansion.

In Ref. [14] the spectrum and the properties of $Q^2\bar{Q}^2$ states have been studied in the MIT bag model.
The $Q^2\bar{Q}^2$ states studied in Ref. [14] have been successfully applied to study the reactions $\gamma\gamma\rightarrow\rho^0\rho^0$
and $\gamma\gamma\rightarrow\rho^+\rho^-$ [16]. 
Recently, a $0^{++}$ resonance, $f_0(1810)$, has been discovered in $J/\psi\rightarrow\gamma\omega\phi$ [17]. If this new resonance is just one of the ordinary mesons
the process $J/\psi\rightarrow\gamma f_0(1810), f_0(1810)\rightarrow\omega\phi$ would be doubled OZI suppressed. In Ref. [18] the new $\omega\phi$ resonance
has been interpreted as the production of the $Q^2\bar{Q}^2$ state, $C^s(\underline{9})$ [14], in $J/\psi$ radiative decay and the doubled OZI suppression is avoided. 

Now we take the $0^{++}$ $C^{ss}(36)$ state studied in Ref. [14] as  
the $s\bar{s}s\bar{s}$ state
\begin{equation}
C^{ss}(36) = \eta_s \eta_s,\;\eta_s = s\bar{s},
\end{equation}
whose mass has been predicted to be 
\begin{equation}
m = 1950 \textrm{MeV}.
\end{equation}
This value is very close to the mass of the $f_0(2100)$. 
The color wave function of $C^{ss}(36)$ state consists of color octet-
color octet and color singlet-color singlet two parts [14]. The recoupling coefficients of this state
are shown in Table I [14]( the color octet is indicated by underline).
\begin{table}[h]
\begin{center}
\caption {Table I Recoupling Coefficients}
\begin{tabular}{|c|c|c|c|c|} \hline
    & PP  &  VV & $\underline{P}\cdot\underline{P}$&$ \underline{V}\cdot\underline{V}$  \\ \hline
$C^{ss}(36)$ & -0.644    & 0.269 &-0.322 &-0.639 \\ \hline
\end{tabular}
\end{center}
\end{table}

In this study the $C^{ss}(36)$ with mass 2100 $\textrm{MeV}$ is taken as the $f_0(2100)$ and
the decays and productions of the $C^{ss}(36)$ state are investigated. We study the decays first, then the productions.
Through a "fall apart" [14] mechanism the  $C^{ss}(36)$ decays to $\eta\eta,\;\eta\eta',\;\eta'\eta'$, $\phi\phi$, etc..
For the decay $s\bar{s}s\bar{s}\rightarrow PP$ the $s\bar{s}s\bar{s}$ is expressed as
\begin{eqnarray}
s \bar{s} = {1\over\sqrt{3}} \eta' - \sqrt{{2\over3}} \eta,\nonumber \\
s \bar{s} s \bar{s} = {2\over 3} \eta \eta + {1\over 3} \eta' \eta' -{2\sqrt{2}\over 3} \eta \eta',
\end{eqnarray}
where $\eta = {1\over\sqrt{6}}(u \bar{u} + d \bar{d} -2 s \bar{s})$ and $\eta' = {1\over\sqrt{3}}(u \bar{u} + d \bar{d} + s \bar{s})$ are taken. 
The three decay amplitudes are via the mechanism of "fall apart" obtained 
\begin{eqnarray}
<\eta \eta |T| f_0(2100)> = -0.644\; a\; m\; {4\over 3},\nonumber \\
<\eta \eta' |T| f_0(2100)> = 0.644\; a\; m\; {2\sqrt{2}\over 3},\nonumber \\
<\eta' \eta' |T| f_0(2100)> = -0.644\; a\; m\; {2\over 3}
\end{eqnarray}
where a is a unknown constant from the mechanism of the fall apart, -0.644 is the recoupling coefficient from Tab. I,
\(m = 2100 \textrm{MeV}\).
The decay widths are derived as
\begin{eqnarray}
\Gamma(f_0\rightarrow \eta\eta) = {0.644^2\over 18 \pi} a^2 m (1 - {4m^2_\eta\over m^2})^{{1\over 2}} = 1.34\times 10^{-2} a^2 \textrm{GeV},\nonumber \\
\Gamma(f_0\rightarrow \eta\eta') = {0.644^2\over 9\pi} a^2 \{{1\over 4m^2}(m^2 + m^2_\eta -m^2_{\eta'})^2 - m^2_\eta\}^{{1\over2}} = 1.05 \times 10^{-2} a^2 \textrm{GeV},\nonumber \\
\Gamma(f_0\rightarrow \eta'\eta') = {0.644^2\over 72 \pi} a^2 m (1 - {4m^2_{\eta'}\over m^2})^{{1\over 2}} = 1.58\times 10^{-3} a^2 \textrm{GeV}.
\end{eqnarray}
The ratios of the three decay channels of the $f_0(2100)$ are
\begin{equation}
\Gamma(\eta\eta) : \Gamma(\eta\eta') : \Gamma(\eta'\eta') \sim 1 : 0.78 : 0.12.
\end{equation}
This $s\bar{s}s\bar{s}$ scheme [14] predicts very small decay rate for the channel $\eta'\eta'$, which is caused by two factors: small phase space and small 
coefficient of the decomposition (3).   
This mechanism also predicts a smaller decay rate for the channel $\eta\eta'$.  
In Ref. [3] a weaker peak in the channel $\eta\eta'$ at 2150 $\textrm{MeV}$ has been reported.

It is known that there is mixing between $\eta,\;\eta'$ and the $0^{-+}$ glueball which is believed to be $\eta(1405)$ [19].  
The mixing makes the $0^{-+}$ glueball $\eta(1405)$ have $s \bar{s}$ component [19]. Therefore, the decay $f_0(2100)\rightarrow \eta\eta(1405)$
exists and have a small decay rate. 
Using the results presented in the first paper of Ref. [19] it is estimated $\Gamma(f_0(2100)\rightarrow\eta\eta(1405)) \sim 0.09 \Gamma(f_0(2100)\rightarrow\eta\eta)$.
However, this decay channel will provide useful information for accurate determination of the $\eta-\eta'-\eta(1405)$ mixing.  
Because the coefficients of the mixing determined by different authors [19] are different the mixing effects on Eqs. (5,6) won't be studied in this paper.

In the same way the decay $f_0(2100)\rightarrow\phi\phi$ can be studied
\begin{eqnarray}
< \phi_{\lambda_1}(k_1) \phi_{\lambda_2}(k_2) | T | f_0(2100)> = 2\; b\;m\; 0.269\; \epsilon^{\lambda_1}(k_1)\cdot \epsilon^{\lambda_2}(k_2),\nonumber \\
\Gamma(f_0\rightarrow \phi \phi) = {1\over 8\pi} (0.269)^2 m (1- {4m^2_\phi\over m^2})^{{1\over2}}\{2 + {m^4\over 4m^4_{\phi}}(1 - {2 m^2_{\phi}\over m^2})\} b^2 
= 0.633\times 10^{-2}\; b^2 \;\textrm{GeV}
\end{eqnarray}
where b is a parameter and the relationship between a and b is unknown.

Because the parameter b is unknown and it is not able to do reliable prediction for the decay rate of the $\phi\phi$ channel.
It is worth to search for the resonance $f_0(2100)$ in the $\phi\phi$ channel. Comparing with the $\eta\eta$ channel, the small phase space 
and the recoupling coefficient of the $\phi\phi$ channel make the decay rate of $f_0(2100)\rightarrow\phi\phi$ much small. However, there are other two factors
which enhance the decay rate of $f_0(2100)\rightarrow\phi\phi$. The amplitude of this decay contains a factor $\epsilon^{\lambda_1}(k_1)\cdot \epsilon^{\lambda_2}(k_2)$,
where $\lambda_i(i=1,2)$ and $k_i(i=1,2)$ are the polarization and momentum of the two $\phi$ mesons respectively. This polarization factor contributes a factor 4.38 to the
decay rate. The second factor is that for the channel $\phi\phi$ the factor for the $s\bar{s}s\bar{s}$ component is one and for the $\eta\eta$ channel this factor is ${4\over9}$ (3). 
All the factors together makes the decay rate of  $f_0(2100)\rightarrow\phi\phi$ not too small. If $a \sim b$ is assumed we obtain
\[\Gamma(f_0(2100)\rightarrow\eta\eta) : \Gamma(f_0(2100)\rightarrow\phi\phi) \sim 1 : 0.47.\]
There are experimental study on $J/\psi\rightarrow \gamma \phi \phi$ [20], in which the $0^{++}\;\phi\phi$ at $2100\textrm{MeV}$ is not studied.
It is worth to search for the $f_0(2100)$ in the $J/\psi\rightarrow\gamma\phi\phi$.

In the picture of four quark state the mechanism of the productions of the $\eta\eta$ resonance $f_0(2100)$ ($C^{ss}(36)$) in $p\bar{p}$ collisions and $J/\psi$radiative decay
can be understood qualitatively. A proton is made of uud quarks and the $f_0(2100)$ is a $s\bar{s}s\bar{s}$ state. How this $s\bar{s}s\bar{s}$ state is
produced in $p\bar{p}$ collision ? 
It is known that half of the energy of a proton is carried by gluons. Therefore, $p \bar{p}\rightarrow g g +..., g g\rightarrow f_0(2100)$ is the process
for the production of the $f_0(2100)$ 
in $p \bar{p}$ collisions. In QCD the $J/\psi$ radiative decay is described as $J/\psi\rightarrow \gamma g g$.   
Therefore, the same $g g \rightarrow f_0(2100) \rightarrow ...$ is responsible for the production of the $f_0(2100)$ in $J/\psi$ radiative decay. 
In this study the $f_0(2100)$ is taken as the 
$C^{ss}(36)$ in which there is a component -0.639  $\underline{\phi}\cdot\underline{\phi}$ (Tab. I), where $\underline{\phi}$ is the color octet $\phi$.
The Vector Meson Dominance (VMD) works well in particle physics, in which photon is coupled to the vector mesons ($\rho,\;\omega,\;\phi$). A similar mechanism is proposed 
in Refs. [21,22], in which gluon is coupled to color octet vector, $\underline{V}$
\begin{equation}
{1\over\sqrt{2}} g_s\; g_{\underline{\phi}}\; g^a \;\underline{\phi}^a,
\end{equation}
where the a is the color index, $g_s$ is the coupling constant of QCD, $g^2_{\underline{\phi}} = {2\over3} g^2$, $g = 0.395$ is determined in Ref. [15], $g^a$ is the
gluon field.        
The process $g +g \rightarrow \underline{\phi} \underline{\phi} \rightarrow f_0(2100) \rightarrow \eta \eta$ is responsible for the productions of the $f_0(2100)$
in both $p \bar{p} $ annihilation and $J/\psi$ radiative decay. 
Using this mechanism, the amplitude of the production of $f_0(2100)$ is at $O(g^2_s)$. Comparing with glueball production in $J/\psi$ radiative decay, the amplitude of the 
production of the $f_0(2100)$ is suppressed by $O({1\over N_C})$ and is at the same order of magnitude of the production of the hadrons made of quarks. 
Roughly speaking, the production rate of glueball in $J/\psi$ radiative decay is at about $O(10^{-3})$ and the $f_0(2100)$ is at $O(10^{-4})$. This is consistent with the analysis
above. 
This mechanism has been applied to study $J/\psi\rightarrow\gamma X(1810), X(1810)\rightarrow \omega \phi$ [18].
Usually, the $J/\psi\rightarrow\gamma \omega \phi$ is a double OZI suppressed process. However, if X(1810) is a four quark state the double OZI suppression no longer exists.

In summary, the study shows that all the decay properties of the $\eta\eta$ resonance $f_0(2100)$ can be understood by the four quark state $s\bar{s}s\bar{s}$, $C^{ss}(36)$.
The $\eta\eta$ channel is the dominant decay mode. The study predicts that the $f_0(2100)\rightarrow\eta\eta'$ has less decay rate. The decay channels, $\pi\pi,\; K\bar{K},\;
4\pi$, are at next leading order in $N_C$ expansion and suppressed. The existence of the decay mode of $f_0(2100)\rightarrow\phi\phi$ is predicted and the measurement of this channel
is significant for the $q^2\bar{q}^2$ scheme of the $f_0(2100)$.  
Because of $\eta-\eta'-\eta(1405)$ mixing the $\eta(1405)$ should have small $s\bar{s}$ component. Therefore, $f_0(2100)\rightarrow\eta\eta(1405)$
exists. The production of $f_0(2100)$ in $J/\psi$ radiative decay and $p\bar{p}$ collisions are resulted in $g g\rightarrow f_0(2100)$.

\end{document}